\def \yskip{\penalty-50\vskip3pt plus 3pt minus 2pt}
\def \abc#1#2#3#4 {\reference#1, {\sl#2}, {\bf#3}, #4}
\def \blank {\lower 5pt\hbox to 0.75in{\hrulefill}}
\def \kms{\rm{km}~$\rm{s}^{-1}$}
\def \cc{$\rm{cm}^{-3}$}
\def \msyr{$M_{\odot}~yr^{-1}$}
\def \cm{~\rm{cm}}
\def \s{~\rm{s}}
\def \km{~\rm{km}}
\def \kms{$~\rm{km}~{\rm s}^{-1}$}
\def \gm{\rm{gm}}
\def \K{~\rm{K}}
\def \g{~\rm{g}}
\def \AU{~\rm{AU}}
\def \ergs{~\rm{ergs}}
\def \erg{~\rm{erg}}
\def \days{~\rm{days}}
\def \yrs{~\rm{yrs}}
\def \yr{~\rm{yr}}
\def \day{~\rm{day}}
\def \pc{~\rm{pc}}
\def \kpc{~\rm{kpc}}
\documentclass[12pt,preprint]{aastex}

\begin{document}
\small
\setcounter{page}{1}

\begin{center}
{\bf 
A NEW LOOK AT THE EVOLUTION OF WOLF-RAYET \\
CENTRAL STARS OF PLANETARY NEBULA }
\end{center}
\begin{center}
Orsola De Marco \\
Department of Astrophysics, American Museum of Natural History \\
Central Park West at 79th Street, New York, NY 10024 \\
orsola@amnh.org \\
\bigskip
and \\
Noam Soker\\
Department of Astronomy, University of Virginia \\
on sabbatical from 
Department of Physics, University of Haifa at Oranim\\
Oranim, Tivon 36006, ISRAEL; soker@physics.technion.ac.il
\end{center}
\begin{center}
On the basis of recent observational evidence and new theoretical results,
we construct a speculative scenario for the evolution of Wolf-Rayet 
central stars of planetary nebula. Although single star evolutionary
calculations have recently succeeded in reproducing the composition
of these objects, it is clear from the latest infra-red observations
that a new perspective has to be adopted: the simultaneous
presence of carbon- and oxygen-rich dust (double dust chemistry), while being a rare phenomenon 
for H-rich central stars, is found around the vast majority of 
cool Wolf-Rayet central stars. This correlation between Wolf-Rayet 
characteristics and double dust chemistry
points to a common mechanism. Within the binary evolution framework established by
Soker,
two scenarios are proposed, responsible for the majority (80-85\%) and
minority (15-20\%) of Wolf-Rayet central stars. 
In the first scenario, proposed here for the first time,
a low mass main sequence star, brown dwarf or planet spirals into the 
Asymptotic Giant Branch star inducing extra mixing, hence a chemistry change, and 
terminating the Asymptotic Giant Branch evolution. In the
second scenario, previously proposed, a close binary companion is responsible for the formation
of a disk around either the binary or the companion. This long-lived disk harbors the
O-rich dust. Both models are {\it speculative}, although supported by 
several observations and recent theoretical results. 
\end{center}


{\it Subject headings:} planetary nebulae$-$stars: mass-loss $-$ stars:
abundances $-$ stars:Wolf-Rayet $-$ planetary systems $-$ Stars:low mass, 
brown dwarfs

\section{INTRODUCTION}
\label{sec:introduction}

Wolf-Rayet central stars (also referred to as [WC] central stars (CS), to differentiate 
them from massive Wolf-Rayet stars; van der Hucht et al. 1981) 
of planetary nebulae (PNe) comprise $\sim 7 \%$ of the entire 
CS population (Tylenda 1996).  
Together with PG1159 CSs, their descendants (Werner 2001), 
we take the H-deficient
CS population 
to be $\sim 9\%$\footnote{Based on the relative numbers of
known PG1159 stars compared to [WC] CSs (Jeffery et al. 1996), and on the 
[WC] CS statistics of Tylenda (1996), we can determine that the H-deficient
evolutionary channel, excluding the weak emission lines stars (38 objects [Tylenda et al. 
1993] which possibly make up a heterogeneous group), is responsible for $\sim$9\% of all CSs.}.

Before 1997 it appeared that the only
evolutionary scenario able to eliminate hydrogen from the envelope of an
Asymptotic Giant Branch (AGB) star was the {\it born-again} scenario 
(Sch\"onberner 1979; Iben et al.\ 1983). 
According to this scenario, an evolved CS already on the white dwarf
(WD) cooling track can experience a last helium shell flash, briefly 
return to the AGB, and then evolve once again along the
CS horizontal track toward the WD domain. This scenario is observationally
confirmed by several objects (V605 Aql, Clayton \& De Marco 1997; 
Sakurai's object [V4334 Sgr], Nakano et al. 1996, Duerbeck \& Benetti 1996; 
FG Sagittae, Gonzalez et al. 1998;
A30 and A78, Jacoby \& Ford 1983), but is now not thought to be the sole channel
for H-deficient central star evolution. 

Since 1997 a series of theoretical papers (Herwig et al. 1997, 1999,
Herwig 2000, 2001, Bl\"ocker 2001) showed that H-deficiency, 
together with carbon and oxygen
over-abundances, like those determined for [WC] CSs by, e.g., Koesterke
\& Hamann (1997) and De Marco \& Crowther (1998,1999), could be 
achieved by two additional scenarios: 
(i) a late thermal pulse (LTP) during the post AGB evolution,
but not as late as the WD cooling track, and (ii) a 
thermal pulse at the end of the AGB evolution (AGB final thermal pulse or
AFTP), when the envelope mass is already low enough that hydrogen can be
diluted down (in this case the post-AGB CS retains some hydrogen).
With this, the interpretation of [WC] CSs seemed to be nearing a
conclusion. 

Soon thereafter, Infrared Spectrographic Observatory (ISO) observations 
of [WC] CSs, showed C-rich dust (in the form of polycyclic aromatic 
hydrocarbons [PAH]) nested inside a shell of hot O-rich dust 
(in the form of amorphous and/or crystalline silicates). 
Both dust phases are hot, indicating
proximity to the hot luminous star (e.g. Waters et al. 1998a, Cohen et al. 1999). 
In the light of a single star scenario, this points to a recent ($\sim$1000~yr)
transition
between the O- and the C-rich AGB phases. In view of the fact that,
according to the same scenario, the transition can happen at any time during the
thermally pulsating AGB evolution (lasting 500\,000 to 10$^6$ years; Vassiliadis \& Wood 1994),
a transition right at the end of the AGB evolution should be a rare occurrence.
The fact that 50\%\footnote{This number is determined from ISO observations of 
16 [WC] CSs, 10 cool ones of which 7 have double dust
chemistry (Waters et al. 1998a [BD+30$^{\rm o}$3639, He2-113], 
Cohen et al. 1999 [CPD-56$^{\rm o}$8032], Cohen 2001 [IRAS07027-7934], Hony
priv. comm. [SwSt1, He2-142 and He2-459]; M4-18, K2-16 and
NGC40 do not [Hony priv. comm, Hony, Waters \& Tielens  2002]), and 6 hot ones, of which one has
double dust chemistry (NGC5315; Hony, priv. comm.). We note also that of the
6 [WCE] CSs, only 3 have high quality data, of which one has double dust
chemistry and 2 have only PAH features.} 
of the [WC] CSs have the double chemistry (versus only
6\% in the H-rich CS group\footnote{For this estimate we counted
3 PNe with double dust chemistry
and H-rich CSs, namely
He2-138, He2-131 (Barlow, priv. comm.) and M3-44 (Hony, priv. comm.). Two
additional PNe with double dust chemistry, have undetected CSs 
(NGC6302 and NGC6537) are were not included in our statistics.}), therefore points to an
event that caused the chemistry change as well as subsequent H-deficiency,
all within a short time. 

In this paper we examine evolutionary scenarios which might naturally lead to 
the double dust chemistry {\it as well} as the [WC] stellar composition. 
This {\it speculative} analysis is developed within the framework of binary 
evolution, including planetary companions, as developed by Soker (1997),
and takes into account the available observations of CSs and their PNe. 
The current paper is also motivated by a recent paper by 
Rosner et al.\ (2001), who introduced a new dredge-up mechanism 
by shear instabilities, which can 
dredge-up carbon and oxygen from WDs
before nova outbursts; we propose that such a process can turn an 
O-rich AGB star into a carbon star and then into a [WC] CS in quick
succession. 
We discuss a plausible origin for the shear mixing within
the framework of accretion disks around cores of some AGB stars. 
In $\S 2$, we review the need for a new scenario. 
In $\S 3$, we list the observational results on which we base our
proposed scenario, which is described in $\S 4$, together with
some predictions.
Our main results are summarized in $\S 5$. 

\section{THE FAILURE OF EXISTING SCENARIOS FOR THE EVOLUTION OF
[WC] CENTRAL STARS}
\label{sec:failure}

According to the sources listed in Note~2, 
ISO spectra (2-160 $\mu$m) revealed the presence of hot amorphous and/or
crystalline silicates 
(O-rich) dust around 80\% of the observed cool [WC] CSs 
([WCL] - {\it L} for late; Cohen 2001). 
Three stars (M4-18, K2-16 and NGC40; Hony et al. 2002, Hony prov. comm.) did not show the silicates. 
The double dust chemistry was detected around 1 of the 6 observed hot 
[WC] CSs 
([WCE] - {\it E} for early), although three of them do not have high quality data. 
Overall double chemistry is observed around 50\%
of all [WC] CSs. 
This is in stark contrast with the statistics for the H-rich group: 
Among 50 H-rich CSs observed by ISO only 3 (6\%) exhibit the double chemistry
(See Note 3). Below we detail why the single star scenario
cannot explain this fact, and we describe two alternative scenarios that have been proposed,
which however are also not fully in line with the observations.

\subsection{The single-star scenario}
\label{ssec:single}

The high incidence of double dust chemistry encountered in the [WC]
CS group is extremely peculiar from a statistical point of view.
About 70\% of all AGB stars undergo enough third dredge-up to rise
their atmospheric C/O ratio above unity (as revealed by the C/O ratio in the PN;
Kingsburgh \& Barlow 1994). Until then, the O-rich AGB
star (C/O$<$1) condenses O-rich dust in its ejecta, since all of the C atoms are locked into
CO molecules, leaving the excess O atoms to form silicates.
When the third dredge-up rises the C/O ratio above unity the AGB star
starts to produce C-rich dust. From theory and observations we expect only a small minority 
of the stars at the end of their AGB evolution to have only 
{\it recently} 
transited to a C-rich
chemistry: Either they have not transited at all (30\% of the cases) or they have
mostly done so hundreds of thousands of years ago (since the thermally-pulsing AGB
evolution, when the third dredge-up can happen, lasts ~500\,000-1$\times$10$^6$~yr 
and comprises between 5 and 25 thermal pulses,
see, e.g., the calculations of Vassiliadis \& Wood 1994).

Since the O-rich dust from the O-rich AGB phase
(C/O$<$1) is still visible
(i.e. hot) around 80\% of the [WCL] CSs and 50\% of all [WC] CSs, the single star scenario implies that
the transition between the two regimes must have
happened recently (as early as 1000 years in the case of CPD-56$^{\rm o}$8032 [Cohen et al. 1999], He2-113 and
BD+30$^{\rm o}$3639 [Waters et al. 1998a]). 
However, from the considerations above, there is no reason why the [WC] CSs
should preferentially exhibit the 
theoretically rare double dust chemistry phenomenon. On the contrary,
since the [WC] CSs are the most extreme case of dredge-up (with stellar 
C/O ratios close to the intershell value), we should be surprised that, as might be indicated by the
double dust chemistry, the vast majority of them
were O-rich till the very end of their AGB life.
(This is unlikely unless, by an
improbable fine-tuning, a first thermal pulse to raise
the C/O ratio above one, results in just the right opacity 
increase 
to expel the envelope in a short time 
and terminate the AGB evolution
[Sweigart 1999]).
The LTP and AFTP cases in Herwig's (2000) single star evolution calculation
do transform the H-rich AGB star into an H-poor one but {\it do not
preferentially} happen in the life of O-rich AGB stars.

\subsection{The disk-storage scenario}
\label{ssec:disk}

In an attempt to explain the high incidence of double dust chemistry around [WC] stars
several authors envisaged a way of storing O-rich dust in a circumstellar disk.
In this way the constraint of a recent AGB transition from O- to C-rich chemistry
is relaxed, since the O-rich dust can reside and be stable in the disk for
a long time.
In this model the silicate dust from the O-rich AGB phase is stored  
in a Keplerian disk, most likely either around a close binary system 
(e.g., Waters, Trams, Waelkens 1992; Jura, Balm, \& Kahane 1995; Waters 
et al.\ 1998b; Van Winckel et al.\ 1998; Jura, \& Kahane 1999; 
Jura, Chen, \& Werner 2000; Jura, Webb, \& Kahane 2001;
Jura, Chen, \& Plavchan 2002), 
like in the carbon stars BM Gem and EU And (Barnbaum et al.\ 1991), 
or around the companion (Yamamura et al.\ 2000).
Such binary systems are likely to form bipolar PNe, e.g., the Red Rectangle
(Soker 1997; by bipolar PN we refer to a PN having two lobes with an equatorial waist
between them). 
However, as we elaborate on in the \S \ref{sec:observations}, most [WC] CSs are
in elliptical rather than bipolar PNe. 
Elliptical PNe are more likely to result from stellar or substellar
companions which enter a common envelope (Soker 1997), or wider stellar companions
which blow weak jets (Soker 2001a); in either of these cases we don't expect the formation of
a circumbinary disk. We conclude that the disk-storage model might not be
be viable to explain the double dust chemistry around all the [WC]
CSs (but see also \S 4.3).

\subsection{The Oort cloud scenario}
\label{ssec:oort}
 
A second alternative scenario for the formation of [WC] CSs with
double circumstellar dust chemistry, is that the O-rich dust  
dates back to when the star was on the main sequence, and is stored in a
structure similar 
to our own Oort cloud (Cohen et al. 1999, 2002). 
This idea is very preliminary and presents considerable drawbacks.
As discussed in \S \ref{sec:observations}, 
the hotter [WCE] CSs appear to present the double dust chemistry in fewer
cases (17\%) than the cooler [WCL] CSs (80\%), possibly implying that
the O-rich dust disappears
from the star vicinity in a few thousand years of evolution, as the star gets hotter.
This is contrary to expectations from evaporated Oort cloud comets
which lasts a long time. 
In addition, this scenario requires a massive Oort-type cloud and/or 
Kuiper-like belt since the mass of the O-rich dust is several times 
10$^{-4}$~M$_\odot$ (Cohen et al. 1999), while the Oort cloud and Kuiper belt combined are
one order of magnitude less massive (Stern \& Weissman 2001; Kenyon \& Windhorst 2001). 
Additionally, Molster et al. (2001)
found the degree of crystallization in the O-dust around NGC~6302 
(a PN with double-dust chemistry, but unknown central star type)
to be much higher than what is known
for comets in our own solar system,
pointing to a different origin for its silicate dust. 
Last, it is not clear why Oort-type
clouds should be more common among the progenitor of [WC] CSs. These considerations make the 
Oort cloud scenario unattractive to explain the dust chemistry around [WC] CSs.

\vspace{0.5in}
In conclusion, 
together with the single star scenario, 
the disk storage and the Oort cloud scenarios present considerable
drawbacks. 
We therefore propose a fourth alternative scenario.
In this scenario the emergence of the [WR] CS type (i.e. extreme
hydrogen deficiency), 
has to arise naturally in unison with the double dust chemistry. 
To construct our scenario, we follow the ideas put forward by Soker (1997)
for the evolution of AGB stars in binaries and stars with planetary systems. 
The main binary evolutionary routes of Soker (1997) are summarized in Table~\ref{tab:binary} 
(columns 1-4), where they are divided into primary and secondary scenarios,
responsible for the majority and minority, respectively, of systems in that
morphological class.
The first column gives the morphological class and the estimated fraction
of PNe in the class; the second column, gives the main morphological features,
while the third and fourth columns give the major and minor evolutionary routes
that lead to the formation of PNe in that class according to the binary model.
In the fifth column we describe the probable role of planets, brown dwarfs
and low-mass main sequence stars in shaping the PN morphology. It is this type
of interaction that we propose to be responsible for he majority of [WC] CSs.

\begin{table}
\begin{tabular}{|l|l|l|l|l|}
\hline
Morphology          & Main                 &Primary binary     & Secondary binary  &Possible role of        \\
class (\% of        & morphological        &scenario (A)       & scenario (B)      &planets and low-mass    \\
CS in class)        & characteristic       &                   &                   &companions              \\
\hline
Round               & All parts            &No binary          &  --               & Only very low          \\
(5-10\%)            & are round            &interaction        &                   & mass planets can       \\
                    &                      &                   &                   & enter the envelope     \\
\hline
Moderate            & Smooth elliptical    &Common envelope    & Stellar           & In case (A) Companion  \\
elliptical          & shape - small        &with low mass MS   & companion         & collide with core and  \\
($\sim 50\%$)       & blobs and jets       &stars, brown       & at 30-100 AU      & may cause extra mixing \\
                    & may exist            &dwarfs or planets  &                   & In case (B) close      \\
                    &                      &                   &                   & planets may exist      \\
\hline
Extreme             & Equatorial mass      &Common envelope    & Stellar           & In case (A) no         \\
elliptical          & concentration        &with a massive (M  & companion         & interacting planets but\\
($25-30\%$)         & (ring) but no        &$\ga$0.3M$\odot$)  & at 30-100 AU      & the companion may      \\
                    & large lobes -        &companion          &                   & collide with the core  \\
                    & Small lobes may      &                   &                   & for M$_2\la$0.3~M$_\odot$ -\\
                    & exist                &                   &                   & In case (B) close      \\
                    &                      &                   &                   & planets may exist      \\
\hline
Bipolar             & Two lobes with       &Companion outside  & Common            & No planets are         \\
                    & an equatorial        &the envelope but   & envelope with     & expected to exist      \\
                    & waist                &close enough to    & a massive         & around the primary     \\
                    &                      &accrete from and   & (M$>$0.5M$_\odot$)& star                   \\
                    &                      &deflect the wind   & companion, at a   &                        \\
                    &                      &of the primary     & late AGB phase    &                        \\
\hline
\end{tabular}
\caption{
Summary of the impact of different binary interaction scenarios on the 
morphology of a PN, as presented by Soker (1997; columns 1-4). Column 5: the proposed role
of planets and other low-mass companions on the formation of PNe in different morphological
classes.}
\label{tab:binary}
\end{table}

\section{OBSERVATIONAL FACTS AND THEORETICAL RESULTS THAT LEAD TO THE PROPOSED SCENARIOS}
\label{sec:observations}

In this Section, we present a list of observations, as well as theoretical results 
which set the stage for the proposed scenario.

\begin{enumerate}
\item
{\bf The high incidence double dust chemistry in [WC] CSs}. ISO observations showed that
the simultaneous presence of O- and C-rich dust is almost exclusively associated with the
H-deficient [WC] CS class (as amply detailed 
in \S \ref{sec:failure}). This fact
points to an alternative mechanism, other than thermally-pulsating AGB scenario, for the 
evolution of [WC] CSs.
\item
{\bf Evolution of double dust chemistry.}
There are 10 
[WCL] CSs observed by ISO of which 7 clearly show the double dust chemistry. 
Two have low quality data (M4-18 and K2-16, Cohen 2001), but do not show the silicate features,
and one, the hottest of the group and only 
[WC8] observed, does not show the double dust chemistry (NGC40; Hony, Waters \& Tielens 2001). 
Of the 6 [WCE] observed by ISO, only one shows the double dust chemistry (the [WC4] NGC5315, Hony priv. comm.).
This could be an indication that as the [WCL] stars evolve to hotter temperatures in a few thousand
years (Vassiliadis \& Wood 1994), the dust is destroyed.
If this argument is correct, it is likely
that none of the hot PG1159 stars, the progeny of [WC] CSs, has the double dust chemistry.
Only one such star was observed by ISO (NGC6765; Hony priv. comm.) and it does not show any
type of dust features.
This depletion of dust as the star grows hotter argues against the scenario 
where long-lived O-rich dust comes from evaporation
of Oort cloud comets, or any other remnant from the main sequence phase.   
\item
{\bf The morphology of PNe around [WC] CS.}
From the 18 spherical PNe listed by Soker (1997), only BD+30$^\circ$3639 has
a [WC] CS. However, this PN is not spherical, but rather
extreme elliptical as revealed by HST observations (Latter et al.\ 2000). 
We conclude that no spherical PN has a [WC] CS.
From the 52 bipolar PNe in Soker's (1997) table 3, 1 PN
has a [WC] CS, i.e., $\sim 2 \%$.
This is compatible with only one [WC] CS in the catalog of 43 bipolar PNe
of Corradi \& Schwarz (1995).
From  the 113 extreme elliptical PNe in Soker's (1997) table 4, 
7 have [WC] CSs (including BD+30$^\circ$3639), i.e., 6\%.
From the 275 PNe in Soker's table 5 of moderate elliptical PNe,
20 have [WC] CSs, i.e., $\sim 7 \%$. As is the case for PNe around
H-rich CSs, and possibly with an additional enhancement, 
there is a tendency for [WC] CSs to reside in elliptical PNe,
but not in spherical or bipolar PNe. Therefore, evolutionary
scenarios resulting in elliptical PNe (in the framework of Soker [1997]),
should be chosen in preference to those that result in spherical or
bipolar ones (i.e., close binaries and single stars).
\item {\bf Lack of close binary nuclei in the [WC] CS population.} 
None of the 16 PNe known to contain close binary nuclei (Bond 2000)
has a [WC] CS; these systems evolved through a common envelope phase, 
in which the companion spins up the AGB envelope (Bond \& Livio 1990).
The lack of [WC] CSs in short period binaries tends to exclude a scenario based on 
common envelope evolution, where the companion survives to the PN 
stage (Tylenda 1996, Tylenda \& Gorny 1993). 
This observational result suggests also that 
envelope rotation is not important. 
\item
{\bf J-type carbon stars.}
J-type carbon stars, which amount to $\sim 15 \%$ of carbon stars,
have $^{12}$C$/^{13}$C$<15$ (Abia \& Isern 2000).
The chemical characteristics of these stars suggests
``... the existence of an extra mixing mechanism that transports material from 
the convective envelope down..." (Abia \& Isern 2000).
Such extra mixing was modeled by Wasserburg, Boothroyd, \& Sackmann (1995).  
All known carbon stars which exhibit silicate dust are J-type (Yamamura et al.\ 2000).
Some are known to have close companions outside the AGB envelope (i.e., at a distance 
greater than about 1 AU), 
e.g., BM Gem and EU And (Barnbaum et al.\ 1991) and V778 Cygni
(Yamamura et al.\ 2000).
Yamamura et al.\ (2000) note also that the correlation between J-type
and silicate dust implies a close relation between the evolution 
at and near the AGB core, and the formation of a long-lived disk outside
the envelope. 
We note the coincidence of the incidence of J-type stars within the C-star group
(15\%) and the incidence of bipolar PNe around [WC] CSs (15-20\%, like for the
H-rich group). If [WC] CSs derive from carbon stars, and if the
J-type stars are always in binary we could speculate that
J-type stars, with close binary companions which 
avoid the common envelope phase, will eventually form PNe with [WC] CSs,
double dust chemistry and bipolar PNe.
Therefore, following the PN morphology statistics, and the incidence of J-type 
stars within the C-star sample, 
we estimate that $\sim 15-20 \%$ of all [WC] CSs are due to wide binary systems, e.g.,
CPD-56$^{\rm o}$8032 ($\S 4.3$).
In \S \ref{ssec:spiraling}, we propose a scenario for the other $\sim 80-85 \%$ of
all PNe. 

\item
{\bf Extra-mixing in novae.} 
The enrichment of CNO nuclei in nova ejecta might require
some `extra-mixing' between the inner CO-rich region of the WD and 
the material above it (Rosner et al.\ 2001, and references therein).
 Rosner et al.\ (2001) argue that this ``dredge-up of C and O'' 
is a result of shear instabilities, resulting from resonant interaction
between gravity waves and large-scale flow of the accreted envelope of
the WD in the nova system. 
 This mixing can occur, for their case, on a time scale of $\sim 5 $ years.
 However, as stressed by Rosner et al.\ (2001), there is a need to 
verify their estimates with numerical simulations. 
We suggest that the same `extra-mixing' process occurs in the progenitors
of some [WC] CSs, whose cores are not dissimilar to the WD in the case of Rosner et al. (2001).
 In our proposed scenario the mechanism may be even more 
efficient, because the mass flow rate into the disk is much higher 
than in cataclysmic variables (\S \ref{ssec:spiraling}), and the CS is expected 
to rotate very slowly, hence a large velocity shear between the 
assumed accretion disk and the core might be present. 
\end{enumerate}

\section{THE PROPOSED SCENARIOS}
\label{sec:proposed}

In this Section we discuss the scenarios which might be responsible for 
the evolution of most (80-85\%) [WC] CSs. The main scenario (\S \ref{ssec:spiraling})
envisages a
low mass companion spiraling into the AGB star and inducing extra mixing 
{\it and} departure from the AGB. The second scenario (\S \ref{ssec:cpd}), is the
disk storage scenario of \S \ref{ssec:disk}, and might be 
responsible for a subgroup (10-15\%) of the [WC] CS group, i.e. the [WC10] CPD-56$^{\rm o}$8032
and those [WC] PNe with bipolar morphology.
The proposed models are summarized in Table~\ref{tab:compare}. 
 
\subsection{The spiraling-in companion}
\label{ssec:spiraling}

The main requirement of {\it any} [WC] CS scenario is to 
turn an O-rich AGB star into a C-rich star and then into a
hydrogen deficient {\it post}-AGB star, i.e., a [WC] CS, in a short time.
In this way,
the O-rich dust, produced during the O-rich AGB phase, 
still resides in the [WC] CS's vicinity.
We envisage that this sequence of events can be triggered by two processes:
(i) A low mass companion to the AGB star that deposits gravitational energy and 
angular momentum as the AGB star expands past its orbit and the companion  enters its envelope,
spiraling in. This results in an enhanced mass-loss rate of the AGB primary. 
Such a binary interaction may lead to slower wind velocity, and even cause
some material to flow back onto the central star (Soker 2001b; see also later). 
(ii) Extra mixing of the intershell material with envelope material.
Only in this way can we generate [WC] CSs, which have the C/O ratio
characteristic of the intershell material (Herwig 2000, 2001) as observed in
[WC] CSs (De Marco \& Barlow 2001).
We now turn to the detailed discussion of points (i) and (ii).

{\bf (i) The increase in AGB mass-loss rate as the companion spirals in 
within the AGB envelope.} The companion will spiral into the AGB star core
within a short time, once the AGB envelope radius reaches the companion's
orbital radius. 
Due to tidal effects, massive planets at a
distance of 3-4 times the AGB stellar radius can spiral in. Since the AGB
radius can reach up to 3-4 AU (e.g. Herwig 2000, where the dredge-up overshoot and
increased opacities result in AGB stellar radii during the thermal
pulses of up to $\sim$5~AU, compared to the $\sim$2~AU reached in the models of Bl\"ocker [1995])
massive planets found at up to $\sim$10 AU
have the potential of being engulfed and spiraling into the AGB star.

Even if the envelope is not entirely expelled by the action of the companion,
the energy released as its mass spirals toward the
core to form the accretion disk (see [ii]), can increase the luminosity and radius of the AGB star
and enhance its mass-loss rate (Siess \& Livio 1999).
The rotating envelope may also have enhanced mass-loss rate. 
Therefore the entire H-rich envelope might be lost in a relatively 
short time after the destruction of the companion.

If the entire envelope is expelled as O-rich gas, i.e., before dredge-up enhancement
is induced (see [ii]), the
accretion disk contains the material of the companion, which is then
mixed with the carbon dredged-up from the He-burning shell.
In BD+30$^{\rm o}$3639, Waters et al.\ (1998a) deduce the mass in the inner C-rich 
dust shell to be $\sim 0.01M_\odot$, while the outer O-rich shell 
contains $\sim 0.37 M_\odot$.
They point out that the mass of the C-enriched shell is the same order of 
magnitude as the sum of the H and He layers in young post-AGB stars.
The low mass of the C-rich shell in BD+30$^{\rm o}$3639 requires the entire
O-rich envelope to be removed before dredge-up. 
We attribute both processes, removal of the O-rich envelope and
the dredge-up, to the destruction of a spiraling-in low mass object,
that was absorbed into the C-rich shell (see [ii]). 

{\bf (ii) The formation of a disk around the core of the AGB star, which 
can induce extra dredge-up from the intershell region.}
In a scenario proposed for novae progenitors by Rosner et al. (2001)
the accretion disk spins up the WD surface layer and induces extra dredge-up
from shear forces. This is proposed as an explanation for the C and O
enrichment in nova ejecta.  
To invoke a similar mechanism for the 
progenitors of [WC] CSs, an accretion disk between the H-burning shell and the AGB envelope
is needed.  
It is not enough to have a close companion to the AGB star, since in most cases close
companions to central stars of PNe do not overflow their Roche lobes,
contrary to the case in novae binary systems, and therefore will not create an accretion disk. 
In addition, observationally, there may be evidence that these systems don't form
[WC] CSs, as we do not observe any [WC] CS in the close binary sample of Bond (2000). 
Instead, we suggest that in most cases the accretion disk is formed 
by {\it destruction} of a low mass object, i.e., a low-mass main sequence star, 
a brown dwarf or a planet.
 The destruction of low mass objects close to the core (by which we mean close
to the H-burning shell, below the H-rich convective envelope) of
AGB stars was studied before.
 Harpaz \& Soker (1994) and Siess \& Livio (1999) studied the response of 
the AGB envelope to a companion's mass, $\sim 0.001-0.01 M_\odot$ ($\sim$1~M$_{\rm J}-10$~M$_{\rm J}$,
where M$_{\rm J}$ is the Jupiter mass), 
which is added to the radiative zone above the H-burning shell and below the convective 
envelope, $\sim 1 R_\odot$ from the core.
Siess \& Livio (1999) found that in the high accretion rate, the bottom 
of the envelope convective region penetrates down into the burning shells, 
leading to hot bottom burning. 
The penetration of the convective envelope zone down into the burning shells 
by itself may cause dredge-up, as noted by Siess \& Livio.
In our scenario, this process can act to enhance the efficiency of
the main mechanism, i.e. shear mixing (see later). 

Soker (1996) proposed that the destruction of brown dwarfs and massive planets 
close to the core of AGB stars may form an accretion disk, 
which may blow two jets. These jets may form fast low-ionization emission 
regions (FLIERs; e.g. Balick et al. 1993), along the symmetry axis of elliptical 
PNe. 
 The accretion disk is similar to that in nova progenitors, but expected 
to be formed from a higher mass accretion rate $\sim 10^{-4} M_\odot \yr^{-1}$
(Harpaz \& Soker 1994; Siess \& Livio 1999), and be more massive (e.g., $\sim$10$^{-4}$~M$_\odot$
compared to $\sim$10$^{-10}$~M$_\odot$ for a nova progenitor [this is a crude estimate, based on
accretion rates, disk energetics and timesales - see Warner 1995]).
 The destruction process, and the life span of the disk, last long enough,
$\sim 100 \yr$, to cause substantial mixing if the time 
scale of $\sim 5 \yr$ found by Rosner et al.\ (2001) 
is applicable, as we assume here. 
The signature of a sole low mass companion (no other
close massive companion is present) that was destroyed near the 
core may be an elliptical PN with jets and/or FLIERs, and possibly 
with a spherical halo (Soker 1996).
 Soker lists (his table 1) 15 such PNe. 
 From these 15, 3 have central [WC] CSs (NGC 6751; NGC 6905; NGC 40)
and one is a WELS (NGC 6543). 
(The presence of H-rich CSs in the above-mentioned morphological class, 
shows that if our proposed scenario is correct, then not all
destroyed low mass stars induce enough mixing and/or that most
PNe in the list of Soker (1996) owe their elliptical shape to another
process, e.g., the companion was destroyed away from the core without 
forming an accretion disk, the destruction rate was slow,
or a wider stellar companion may have shaped the nebula [Soker 2001a]). 

To form an accretion disk, the low mass companion dense gas {\it must} spiral 
all the way to the core vicinity, $\lesssim 0.1 R_\odot$.
When reaching a distance of $\sim 1 R_\odot$, low mass companions are
tidally destroyed (e.g., Soker 1996; Siess \& Livio 1999). 
A Jupiter type planet of mass $M_p$ will be destroyed earlier by 
evaporation (Siess \& Livio 1999), when it gets to a location inside 
the envelope where the envelope temperature exceeds its virial temperature 
of $T_V \sim 2 \times 10^5 (M_p/M_J) \K$ (Siess \& Livio 1999). 
The location in the AGB envelope at which the temperature is equal to
$\sim 2 \times 10^5 \K$ is $r \sim 10 R_\odot$ (Soker \& Harpaz 1999). 
Hence, a Jupiter-like planet will start to be destroyed at
$\sim 10 R_\odot$ from the AGB core.
Even if all the AGB luminosity goes to heat the planet's material to
its virial temperature, the process will take a time of 
$\tau_{\rm eva} \sim 0.1 (M_p/M_J)^2(L_\ast/10^4 L_\odot)^{-1} \yr$
where $L_\ast$ is the AGB stellar luminosity (Harpaz \& Soker 1994;
Siess \& Livio 1999).
Since not all the luminosity is intercepted by the evaporating planet,
the evaporation time will be longer.
The planet, and its dense evaporated material, before becoming hot, 
continue to spiral-in while suffering a strong friction,
and collide with the core within a few orbital periods.
The orbital period at $\sim 10 R_\odot$ is $\sim 0.01 \yr$. 
We therefore conclude that planets having a mass of $M_p \gtrsim M_J$
may survive evaporation till they reach the core of the AGB star, where 
they
may form disks after being tidally destroyed.
Dredge up as a result of a companion colliding with the core of a
red supergiant during a common envelope evolution was discussed 
by Ivanova \& Podsiadlowski (2002).
However, they discuss a massive main-sequence companion rather
that a low mass companion, and did not consider a disk.

The upper limit on the mass of a companion that can create a disk
around the AGB core, is set 
by the requirement that the companion reaches a radius of $\sim 1 R_\odot$,
where it is being tidally destroyed, before the AGB envelope is
lost, 
i.e., before the end of the AGB evolution. 
 Due to the large theoretical uncertainties, we use the following
 observation. 
The central binary system MT Ser of the PN Abell 41 has an orbital 
period of 0.113 days (Bond 2000), which implies an orbital 
separation of $\sim 1 R_\odot$.
 Since the companion was not tidally destroyed, its mass must be 
$M_2 >  0.01 M_\odot$ (Soker 1996, Eq. [3]); more likely it is a main 
sequence star with $M_2 \gtrsim  0.1 M_\odot$.
In most cases more massive secondaries are likely to expel the AGB
envelope before colliding with the core. 
In summary, we argue that planets, brown dwarfs and low mass main 
sequence stars (up to a mass of $\sim 0.1 M_\odot$) can form disks. 
 
Last, if the spiraling-in occurs at an early AGB phase, an elliptical 
PN, but no [WC] CS, will be formed. The reason is that, at that stage, the
mass separation between the He-burning shell and the surface of the
core is too large (e.g., Bl\"ocker 1995), and mixing may not be efficient.

\subsubsection{A prediction of the ``spiraling-in companion" scenario}
\label{sssec:spiraling}

How many systems are expected to result in [WC] CSs? 
If, as we have postulated, 80-85\% of {\it all} [WC] CSs derive from the
destruction of a low-mass companion, then we have to explain 6-7\%
of all CSs (80-85\% of 9\%) with the action of such companion.
 From the 1,200 main sequence stars discussed by Vogt et al.\ (2002) 
44 planets and 4 brown dwarfs have been detected,
i.e., $3.7 \%$ of the surveyed stars were found to have planets. 
For stars with metallicity above solar, [Fe/H]$>0$, the detection fraction
is $\sim 2.6$ times higher than for the entire sample (Vogt et al.\ 2002),
i.e. $\sim 10 \%$.  
Reid (2002) argues that most stars with [Fe/H]$>0.3$ harbor planets
with $M_ \gtrsim 1M_J$. 
It is quite plausible that most PNe are formed from relatively metal-rich
stars, since most PNe are formed from stars having main sequence
mass of $M_{\rm MS} \gtrsim 1.3 M_\odot$ (Allen, Carigi, \& Peimbert 1998),
which are likely to have higher metallicity compared with lower mass,
hence older on average, stars. 

However, the known extra-solar planets are within a few AU of the parent star
and will become engulfed by it much before 
the AGB phase. These planets do not fit our proposed scenario.
However, according to Lineweaver \& Grether (2002) the detection bias for
planets close to their parent star, means that many more planets should exist
at larger distances, and are yet undetected.
We therefore predict that compared with the present detection fraction
of extra-solar planets, there are many more as yet undetected massive planets, 
brown dwarfs, and low mass main sequence stars, at orbital separations 
of $a \sim 3-10 \AU$,
around the progenitors of PNe (mainly with
main sequence mass of $M_{\rm MS} \gtrsim 1.5 M_\odot$). 
Oppenheimer et al. (2001) carried out a multi-color survey for stellar companions.
Their survey is basically complete for companion masses $\gtrsim$40~M$_J$ between $\sim$40 and $\sim$120 AU.
They detect a very low number of companions, corresponding to a binary fraction of 
only $\sim$1\%. However their survey is highly incomplete for companion masses 
$\lesssim$50~M$_J$ and for distances from the primary star of $\lesssim$10~AU.
Therefore, although instructive (i.e., we should not expect a large number
of low mass companions at $\sim$10~AU - the outer orbital limit of our proposed scenario), this
survey does not answer the question of the binary fraction within the orbital separation limits imposed by
our scenario.
We also should take into account that not all planets at an appropriate distance
to enter the AGB envelope will result in [WC] CSs.
If we quantify that about half of the relevant binary and planetary systems form [WC] CSs created
by the companion-destruction mechanism, 
then we require, hence also predict, the total fraction of stars with such
companions to be, very approximately, $\gtrsim 12 \%$ (6-7\% to account for 
the fraction of [WC] CSs, multiplied by a factor of two to account for those
companions that might not enter the AGB envelope) among this type of main sequence stars.

\begin{table}
\begin{tabular}{|l|l|l|l|}
\hline
&\multicolumn{3}{|c|}{Models for the evolution of [WC] central stars}\\
\hline
Observational       & {\bf Single star model:}& {\bf Proposed model:} & {\bf Secondary model:}     \\
facts               & thermal pulsing   & 80-85\% of [WC] & 15-20\% of [WC] CSs: \\
                    & on the AGB        & CSs: low mass   & close companion  \\
                    &                   & companion-core  & outside AGB    \\
                    &                   & collision       & envelope forms  \\    
                    &                   &                 & O-rich disk                 \\     
\hline
[WC] tend to be in  & Not explained:    & In agreement:     & In agreement with  \\
elliptical rather   & expect a          & mostly elliptical,& a subset of       \\   
than in bipolar PNe & spherical PN      & some bipolar PNe  & [WC] CSs           \\  
\hline
No [WC] in          &  Not explained:   & In agreement: most & In agreement: some  \\
spherical PNe       &  expect a         & PNe are elliptical & PNe are bipolar    \\
                    &  spherical PN     &                    &                   \\
\hline
Double dust         & Not explained:    & In agreement:      & In agreement:     \\
chemistry in        & see \S 2              & fast transition    & O-rich dust is    \\
$[$WC$]$            &                   & from AGB to [WC]   & stored in the     \\
                    &                   & +core envelope     & binary disk       \\
                    &                   & extra mixing       &                   \\
\hline
$[$WC$]$ might not   & In agreement:    & In agreement: the  & In agreement:     \\
have a very close    & no companion     & companion is       & companion is not \\
binary companion     &                  & destroyed          & close           \\
\hline
\end{tabular}
\caption{Summary of the observed characteristics of [WC] CSs and their PNe
and the way they can be explained by
different evolutionary models.
Note that the proposed scenario includes a main scenario (responsible
for 80-85\% of all [WC] CSs; third column) and a secondary scenario
to explain about 15-20\% of the [WC] CSs (fourth column).}
\label{tab:compare}
\end{table}

\subsection{The case of CPD-56$^{\rm o}$8032: a stellar companion outside the envelope}
\label{ssec:cpd}

The O-dust reservoir model described in \S \ref{ssec:disk}, presents some drawbacks, 
in that if it alone were responsible for the evolution of [WC] stars,
then we should observe more bipolar PNe around them. However since 
{\it some} bipolar PNe {\it are} observed, we conclude that this
scenario might be responsible for some [WC] CSs (we tentatively
suggest that 15-20\% of [WC] CSs might be explained in this way, 
following the statistics for their PN morphology).

According to this model, a close to, but
outside, the AGB envelope binary companion shapes the AGB wind and 
forms a reservoir of dust, either as a circumbinary disk 
(e.g., Van Winckel et al.\ 1998; Waters et al.\ 1998b; 
Jura, et al.\ 2001; Jura et al.\ 2002),
or a disk around the companion (Yamamura et al.\ 2000). 
As explained (\S \ref{ssec:disk}), 
this model does away with the need to invoke a fast 
transition from an O-rich AGB star to a [WC] CS since
the O-rich dust-disk can live for more than several thousand years
(H. Van Winckel 2001 priv. comm.). 
The star can then turn into a H-deficient [WC] CS via a thermal pulse on
the post-AGB track, as in Herwig's (2000) single star scenario.

Alternatively, the formation of a disk around the core of the AGB star, leading to
sufficient dredge-up to turn the AGB star into a [WC] CS, may result from
back-flowing material. 
Bujarrabal, Alcolea, \& Neri (1998; see also Soker \& Livio 1994 for
accretion at late a PN phase), motivated to explain jets in PNe,
suggested that back-flowing material, expected to possess large specific 
angular momentum, will spin up the post-AGB envelope and form an accretion
disk around the compact post-AGB star. The spun-up post-AGB envelope is expected to have enhanced mass-loss
rate which may deplete the star of its hydrogen-rich gas.
An accretion disk, if formed, may also cause extra mixing.
Zijlstra et al.\ (2001) suggest that accreted back-flowing material 
during the post-AGB phase may lead to a slower evolution along the 
post-AGB track. 
  Post-AGB stars depleted of refractory elements which
compose the dust particles (e.g., Van Winckel et al.\ 1998)
can be formed by accretion of a dust-depleted circumstellar gas 
(Waters et al. 1992), most likely in binary systems
(Van Winckel 1999). 
 Soker (2001b) studied the condition for the occurrence of a significant
back-flow during the AGB and post-AGB phases, and found that 
for a significant back-flow to occur, a slow dense
flow should exist. 
He further argued that the requirement for both high mass-loss rate per 
unit solid angle and a very slow wind, such that it can be decelerated 
and flow back, probably requires close binary interaction.
 If the back-flowing mass has a large specific angular momentum, as expected 
in strongly interacting binary systems, the back-flowing material may form
an accretion disk around one or two of the two stars.
Some of the back-flowing gas may be accreted around the companion,
and some may be caught in a circular orbit together with the companion.

An example of the disk storage scenario might be CPD-56$^{\rm o}$8032. In this system, Cohen et al. (2002)
observed visible light variability with a 5 year period. If the
periodicity is confirmed by further observations, this might point
to the existence of a dust disk; This might be a circumbinary disk,
with a $\sim$10-yr precession period, or
alternatively, a dusty disk around the eclipsing companion with a period of $\sim 5 \yr$
(Cohen et al. 2002).
A second example might be the the CSs NGC 6302 and NGC 6537 (Molster et al. 2001, 2002). These PNe
have bipolar morphology and therefore are expected to have binary companions
outside the envelope (Soker 1997). We also note that these PNe have undetected CSs,
so that we do not know whether they belong to the [WC] or H-rich class. (The disk
storage scenario could apply to either [WC] or H-rich groups. In the former case,
the timing of the last thermal pulse (LTP or AFTP, as in the single star model) will
turn the primary into a [WC] or back-flowing material will induce extra mixing. In
the latter case, no timing is necessary for the last AGB thermal pulse, and the
back-flowing material, if present, did not result in extra mixing and mass-loss.)

Finally, we note that another possibility is that binary systems
with orbital separation of $\sim 20 -100 \AU$, which are thought to
form mainly elliptical and extreme elliptical PNe, but some bipolar PNe
as well, (Soker 2001a), harbor planets around one or two of the stars.   
The [WC] CS formation mechanism is then as in the previous Section (\S \ref{ssec:spiraling}),
but a bipolar or extreme elliptical PN is formed. This mechanism, too,  could account for 
some of the [WC] CSs in bipolar PNe.
A planet candidate orbiting a star in a binary system, although a wide one
with an orbital separation of $>600 \AU$, was
announced recently by Zucker et al.\ (2002). 
Other systems with both a close planet and a stellar companions at  
$\sim 100 \AU$, are known, and it is likely that there are many more    
such systems (Mazeh \& Zucker 2002).

\section{SUMMARY}
\label{sec:summary}

 The goal of this paper is to present a plausible scenario to
explain [WC] CSs (i.e., Wolf-Rayet central stars of PN).
In building the scenario, we carefully considered some observational
results and processes that were 
proposed in the past for different astrophysical objects and/or to 
account for other properties of PNe (\S \ref{sec:observations}).
In order to account for the presence of O-rich dust around [WC] CSs, 
we proposed that a low mass companion, 
$0.001M_\odot \lesssim M_2 \lesssim 0.1 M_\odot$, i.e. a low-mass 
main sequence star,
a brown dwarf, or a planet, enters a common envelope
phase with the AGB progenitor. 
The companion enhances the mass-loss rate, hence expedites the termination
of the AGB.
It is then destroyed and its dense material collides with the core 
and forms an accretion disk. Like in the nova scenario of Rosner et al. (2001), 
the accretion disk causes extra mixing between core and envelope material, 
turning the star to a C-rich star, and then to a hydrogen deficient one
(which subsequently develops a fast, dense stellar wind and the [WC] spectral signature,
in virtue of its higher opacity). 
The extra mixing of core material, i.e., dredge-up, is due to 
shear instabilities induced by the accretion disk.

 The low mass companion scenario accounts for most, but not all, [WC] CSs,
since it predicts that [WC] CSs reside in elliptical PNe, while some
[WC] CSs are observed in bipolar PNe.
Some [WC] CSs could be formed by 
the influence of a stellar companion close to, but outside, the 
AGB envelope (e.g. van Winckel et al. 1998; Yamamura et al. 2000). 
In this scenario the companion forms a long-lived reservoir of O-rich dust during
the O-rich AGB phase.
 Back-flowing material during the post-AGB phase can influence the
 evolution in several ways.
 First, it can spin-up the post-AGB stellar envelope leading to enhanced
mass-loss, so that the star expels its hydrogen rich envelope.
 Later, the back-flowing gas may be accreted to the core and slow 
the post-AGB evolution down (Zijlstra et al.\ 2001).
The low entropy back-flowing material may form  an accretion disk,
leading to extra mixing. 
 
As stated above, these processes were studied before for different systems. 
In the merging degenerate dwarfs model for R Coronae Borealis stars
(e.g., Iben \& Tutukov 1985),
C- and O-rich material from the outer layers of a C-O-WD is mixed
by shear forces with a disk formed from a destroyed lower mass He-WD. 
Shear instabilities were also proposed to account for the dredge-up of  
core material in WD nova progenitors (Rosner et al.\ 2001 and references therein), 
where the accretion
disk is formed via Roche lobe overflow from a low-mass main sequence star.
To form an accretion disk around the core of an AGB or post-AGB star,
we turned to processes that have been proposed to explain
other properties of PNe and proto-PNe. 

Our qualitative analysis is supported by observations as well as quantitative
calculations carried out in the cited papers. However, only a set of 
3-dimensional models for the parameter space of an AGB star and a low mass
companion (M$\lesssim$0.1~M$_\odot$), can put this idea on solid grounds. 
The calculations carried out
by Sandquist et al. (1998) for AGB stars of 3 and 5~M$_\odot$ colliding with
main sequence stars of 0.4 or 0.6~M$_\odot$ are the closest models to the
proposed configuration. It is clear from the sensitivity of their results to
some of the initial conditions, that their conclusions are not applicable to the
current scenario. It is however instructive to see how 3-dimensional
calculations lead to different results compared to 1- or 2-dimensional ones
(Terman, Taam \& Hernquist 1994).
Additionally, although a model like that of Sandquist et al. (1998) might
provide the answer to the amount of mass-loss generated by spiraling-in
of low-mass companions into AGB stars, an essential ingredient of our speculative
scenario, it is not clear that sufficient resolution
could be achieved to determine the details of the tidal disruption of the planet,
its transformation into a disk around the core, and the subsequent induction
of extra dredge-up by sheer mixing. 

We finally note that, provided that the spiraling-in companion does result in
a significant enhancement of the AGB stellar mass-loss, such that termination of the AGB
phase is guaranteed, the mechanism for the creation of the [WC] CSs might be
able to operate even in the absence of sheer forces. It was pointed out by
Falk Herwig (2000, priv. comm.) that the radius growth at the time of AGB thermal pulses
(also see, e.g., Fig.~5 of Bl\"ocker [1995]) can, in the presence of
overshoot, extend to 4.5-5~AU for a limited amount of time. During this time, it might reach out to
companions that have not yet entered the AGB star and allow them to spiral into the 
core (as we have explained companions which are even farther out than the nominal
stellar radius can be brought in by tidal effects). 
Following the thermal pulse, the post-thermal pulse convection zone
has the appropriate [WC] abundances. Therefore,
removal of the envelope at this critical moment would naturally generate a [WC] CS, in
a similar way to the AGB AFTP model of of Herwig (2001). In fact the help of the
companion, would remove the fine tuning required by the AFTP model, whereby
the AGB departure mysteriously happens right after a thermal pulse. Indeed the pulse
would be the cause of both the increased mass-loss which allows departure from the
AGB (by companion capture) and the dredge-up needed to leave the star
without hydrogen and with the intershell abundances characteristic of a [WC] CS.

Despite the lack of quantitative calculations, we find that our scenario
makes a few strong predictions.
\newline
(1) Most [WC] CSs have no close stellar companions around them,
because they had low-mass companions, which collided with the core and were destroyed.
Some [WC] CSs, though, are formed by other routes, and may have companions
around them, e.g., CPD-56$^{\rm o}$8032 (PN G332.9-09.9; Cohen et al. 2002, De Marco 2002).
Like CPD-56$^{\rm o}$8032, other CSs, in particular those in bipolar PNe, should have
properties that betray the presence of the companion or the dusty disk around it. 
\newline
(2) The fraction of main sequence stars in the mass range 
$1.5 \lesssim M_{\rm MS} \lesssim 3 M_\odot$ which have
companions in the mass and orbital separation ranges 
$0.001 \lesssim M_2 \lesssim 0.1 M_\odot$ and  
$3 \lesssim a \lesssim 10 \AU$, respectively, 
should be $\gtrsim 12 \%$.
This prediction is actually a requirement for the proposed model
to explain most [WC] CSs. 
\newline
(3) If planets reside mainly in planetary systems, then many [WC] CSs
should have planets at large orbital separations around them which survived 
the evolution, i.e., Saturn, Uranus, Neptune-like planets. 
The possibility of detecting such planets in PNe was discussed by
Soker (1999), and around hot WDs by Chu et al.\ (2001). 
\newline
(4) In some cases the material from the destroyed companion can
comprise a significant fraction of the mass of the latest wind or
the envelope of the [WC] CS.
 We foresee the possibility of  peculiar abundances in some
[WC] CSs and/or their PNe, which can't be explained by simple single stellar evolution.
 
\bigskip

{\bf ACKNOWLEDGMENTS:}
We gratefully acknowledge the discussions with Sacha Hony on the nature of
circumstellar dust as well as thank him for providing a list of ISO PNe observations.
We are also grateful to Falk Herwig for providing insightful comments.
We benefited from discussions with Agnes Acker, Martin Cohen, Mike Barlow, Carsten Dominik, 
Slawomir Gorny, Angela Speck, Romuald Tylenda and Hans Van Winckel. 
OD gratefully acknowledges the support of the Isaac Asimov fellowship 
program.
NS was supported by a Celerity Foundation Distinguished Visiting 
Scholar grant at the University of Virginia, and by a grant from the
US-Israel Binational Science Foundation. 



\end{document}